# The electronic structures of $TiO_2/Ti_4O_7$, $Ta_2O_5/TaO_2$ interfaces and the interfacial effects of dopants

H. Li[*][†], Z. Zhang[†], L. Shi

Department of Precision Instrument, Centre for Brain Inspired Computing Research, Tsinghua University, Beijing, 100084

[†]*These authors contribute equally.*
[*]*Corresponding author email: li_huanglong@mail.tsinghua.edu.cn*

**Abstract**

We study the electronic structures of $TiO_2/Ti_4O_7$ and $Ta_2O_5/TaO_2$ interfaces using the screened exchange (sX-LDA) functional. Both of these bilayer structures are useful infrastructures for high performance RRAMs. We find that the system Fermi energies of both interfaces are just above the conduction band edge of the corresponding stoichiometric oxides. According to the charge transition levels of the oxygen vacancies, the oxygen vacancies are stabilized at the -2 charged state in both $Ta_2O_5$ and $TiO_2$. We propose to introduce interfacial dopants to shift the system Fermi energies downward so that the +2 charged oxygen vacancy can be stable, which is important for the controlled resistive switching under the electrical field. Several dipole models are presented to account for the ability of Fermi level shift due to the interfacial dopants.

**Introduction**

Metal oxides have been widely used in the emerging nonvolatile memory technologies like the resistive random access memory (RRAM) [1,2] and the memristor [3-5]. Both of these devices are characterized by the reversible resistance switching (RS) which is commonly due to the growth and rupture of the conducting filaments (CFs) consisting of charged oxygen vacancies (Ov)[6-8]. Numerous reports have been published on metal oxides operating through this RS mechanism like $TiO_2$ [9,10], $Ta_2O_5$ [11], $SrTiO_3$ [12], NiO [13,14], and so on. Among various oxide candidates, $TiO_2$ demonstrates both unipolar [9,10] and bipolar [15,16] RS. Transmission electron microscopy of the unipolar RS $TiO_2$ has suggested that the CFs in unipolar switching are composed of Magneli phase $Ti_4O_7$ [10]. Recently, real time identification of the evolution of $Ti_4O_7$ CFs in $TiO_2$ has been reported [17]. However, the problem of $TiO_2$ is the co-existence of more than one stable substoichiometric Magneli phases which is unfavorable for long device endurance [18]. On the other hand, $Ta_2O_5$ keeps the longest endurance record among these oxides [11] and it operates in a bipolar manner. The long endurance of $Ta_2O_5$ based devices is partly attributed to only one stable substoichiometric phase, Ta, under 1000°C [18] despite of large amount of metastable substoichiometric phases such as $TaO_2$. The in situ observation of CFs in $Ta_2O_5$ has recently been reported [19].

One of the biggest challenges of these oxide based devices is the requirement for the electroforming process [20,21]. This could induce physical damages to the device due to oxygen gas release [22]. Besides, wide variance of device properties depends on the details of the electroforming which are yet to be understood. Therefore, it is essential to achieve electroforming free RS so as to eliminate the device variance for applicable computer circuits [20,21].

Electroforming free RS has been reported in various material architectures such as $Ta_2O_5/HfO_{2-x}$ [23], $Al_2O_3$ [24], $SrTiO_3$ [25], $TiO_2/Ti_4O_7$ [22] and so on. The latter two examples invoke the CF mechanism. The rationale is either using a reducing electrode (Ti) to create sufficient Ov in the oxide or using a thick Ov reservoir for a readily formed conducting path across the bulk region so that only the switching interface near the electrode is kept. The $TiO_2/Ti_4O_7$ bilayer structure is thus useful to achieve reliable devices. $Ta_2O_5/TaO_2$ bilayer structure has the same effect in this sense that record high device performance has been reported based on this structure [11].

Diverse experimental findings of devices showing either the unipolar behavior or the bipolar behavior lead to certain confusion in the essential mechanism of the RS. There has been numerous reports on the unipolar switching in $Pt/TiO_2/Pt$ cell [10]. At the same time, bipolar switching has also been reported in $Pt/TiO_2/TiO_x/Pt$ cell [22]. The conversion from unipolar switching of the $Pt/TaO_x/Pt$ cell to bipolar switching of the $Pt/TaO_x/Ta_2O_5/Pt$ has recently been reported [26] and been explained by the interface-modified random circuit breaker network model [27]. This phenomenological model does acknowledge the importance of the bilayer oxide interface, while understanding the interface from an atomic level is still necessary. The integration of RRAMs or memristors into the passive crossbar array promises the ultimate scaling potential. Recently, complementary RS has been found in the $Ta_2O_5/TaO_2$ bilayer structure [28], which is useful in solving the sneak path problem in the crossbar array. Therefore, it is important to understand the interfacial structures of $TiO_2/Ti_4O_7$ and $Ta_2O_5/TaO_2$. However, very few of such studies have been reported.

Enhanced device performances (e.g., cycling endurance, ON/OFF ratio, switching speed and RS uniformity) can also be achieved through doping cations (e.g., Al [29], Ti [30], Si [31], Li [20]) or anions (e.g., N [32]) into the metal oxide. The role of these additional dopants in RS is mainly understood from the bulk effects like tunable Ov formation energy. A selection rule of dopants from this perspective has been proposed by first principle study [33]. To the best of our knowledge, no study of the interfacial effect of dopants at the interface of $TiO_2/Ti_4O_7$, $Ta_2O_5/TaO_2$ has been reported. In this work, we study the electronic structure of the $TiO_2/Ti_4O_7$ and $Ta_2O_5/TaO_2$ interfaces and the interfacial effects of dopants on the RS process.

**Computational methods**

The density functional theory (DFT) calculations in this work are performed by the Cambridge Serial Total Energy Package (CASTEP) [34,35]. To optimize the atomic structures, the generalized gradient approximation (GGA) [36] is used to present the exchange and correlation effects of the electrons. A plane-wave basis set with cut-off energy 380 eV and ultra-soft pseudo-potentials are used. Convergence tolerance of $1\times10^{-5}$ eV and $5\times10^{-2}$ eV·Å$^{-1}$ is used for the total energy and force, respectively. The spacing of uniformly sampled k points is minimized to $2\pi\cdot0.3$ Å$^{-1}$ for sufficient convergence. Local density functionals such as GGA are known for underestimation of the band gaps of semiconductors or insulators. Hybrid functionals, such as sX-LDA and HSE functionals [37,38], include non-local exchange-correlation explicitly dependent on wave-functions in order to carry the derivative discontinuity of the energy for integer particle numbers, leading to the opening of band gap. We use sX-LDA to calculate the band structures of the pristine $Ta_2O_5/TaO_2$ and $TiO_2/Ti_4O_7$ interfaces and the formation energies of point defects in bulk $Ta_2O_5$ and $TiO_2$. Cutoff energy of 750 eV and norm-conserving pseudo-potentials are used in sX-LDA calculation. Nevertheless, the band offsets of heterostructures provided by GGA are in general qualitatively similar to those by higher level hybrid functionals, while the latters are significantly more time consuming than GGA as the number of atoms increases to tens or hundreds. GGA has been widely used to study the interfacial properties, e.g., interfacial dopant [39] and interlayer [40,41] tunable effective work function (EWF), interfacial defects induced Fermi level pinning [42,43]. In effect, GGA can be considered as a useful rule of thumb in many cases if the trends rather than the absolute values are of interest, as the scope of the study of interfacial dopants in this work, where only GGA functional is used to demonstrate the trends of band offset shift.

**Results and discussion**

### Bulk crystalline structures of $Ta_2O_5$, $TaO_2$, $TiO_2$ and $Ti_4O_7$

The crystalline unit-cell structures and the corresponding GGA band structures of bulk $Ta_2O_5$, $TaO_2$, $TiO_2$ and $Ti_4O_7$ are shown in Fig. 1; for the sake of comparison, respective sX-LDA band structures are also shown. Crystalline $Ta_2O_5$ has a complex layered structure and has only recently been explained in a simple model [44], known as the λ phase. Stability with respect to the amorphous phase has been predicted [44]. It has three oxygen sites: one interlayer and two intralayer. The GGA band gap is 1.8 eV and is improved by sX-LDA functional to 4.0 eV, in agreement with the experimental value [45]. Crystalline $TaO_2$ has a rutile structure and is shown to be metallic by both GGA and sX-LDA functionals, which is consistent with a recent calculation [19] and the XPS observations [19]. $TiO_2$ has two common polymorphs known as the rutile phase and anatase phase. Experimentally, these two samples can be grown by sputtering and plasma-enhanced atomic layer deposition (PEALD) [17], respectively, which show distinctive RS properties: the anatase phase exhibits high variability in the RS parameters and requires an electroforming process while the rutile phase shows higher uniformity and no distinct electroforming process [17]. The differences were attributed to structural compatibility of rutile phase $TiO_2$ with the Magneli phase CF, resulting in easy recovery of the CF without the need of extensive Ov diffusion [17]. Therefore, the rutile phase $TiO_2$ should be more suitable for RS. The GGA band gap of rutile $TiO_2$ is 1.9 eV and sX-LDA functional amends it to 3.0 eV, in agreement with the experimental value [46]. $Ti_4O_7$ is the main composition of the CFs in $TiO_2$ [10]. Despite of the existence of various phases of $Ti_4O_7$ at different temperatures which show different electronic transport properties [47-50], the atomic structures of these phases bear little difference. We optimize the structure by GGA functional from what was reported by Marezio et al [48]. Both of our GGA and sX-LDA band structures show metallic properties of the relaxed $Ti_4O_7$ model structure. It is noteworthy that the Magneli phase can be obtained from the rutile unit cell by proper transformation matric [51]. The two structures bear clear resemblance-that is, the extended Ov restricted on the (121) shear planes of rutile $TiO_2$ result in Magneli phase $Ti_4O_7$.

### Interface structures of $Ta_2O_5/TaO_2$ and $TiO_2/Ti_4O_7$

The interface structure of $Ta_2O_5/TaO_2$ as shown in Fig. 2(a) is built from respective crystalline phases, which significantly reduces the modelling complexity. Although the amorphous structure could be more useful in experiment, it is still advantageous to use the λ $Ta_2O_5$, which has adaptive structure and predicted stability over the amorphous phase [44], to understand the key structures and the energetic features of the interface. The interface is chosen to be perpendicular to both of the (001) direction of $Ta_2O_5$ and the (110) direction of $TaO_2$, which minimizes the lattice and bond mismatch. The six-fold coordination of the interfacial Ta is retained. Layered structure is clearly seen across the interface. Two identical interfaces are included in the supercell model so that any unintentional dipole effect across one interface is cancelled out at the other. We optimize this structure by fixing two lattices parallel to the interface to those of λ $Ta_2O_5$ and allowing only the lattice vertical to relax. We calculate both the GGA and sX-LDA band structures of $Ta_2O_5/TaO_2$ interface, as shown in Fig. 2(b, c). The Fermi energy of the interface lies above the conduction band edge of $Ta_2O_5$.

The interface structure of $TiO_2/Ti_4O_7$ is shown in Fig. 3(a). The interface is nearly strain free and defect free due to structural compatibility. The same geometry optimization strategy as mentioned above is used. From both the GGA and sX-LDA band structures, as shown in Fig. 3(b, c), the Fermi energy of the interface is above the conduction band edge of $TiO_2$, resulting in an ohmic contact. This is in contrast to the GGA+U result of Padilha et al [52], where type-II heterostructure of $TiO_2/Ti_4O_7$ with band gap 0.16 eV was found. Applying +U for states that are more appropriately described as delocalized or itinerant bands such as Ti d states, however, is unwarranted and may lead to spurious results [53]. The discrepancy may invoke the need for an insight from beyond DFT such as GW [54].

### Ov formation energies in $Ta_2O_5$ and $TiO_2$

The results of both $Ta_2O_5/TaO_2$ and $TiO_2/Ti_4O_7$ interfaces have important implications of the stabilized charge state of Ov as will be shown, considering the dependence of the charge state stability of Ov on the position of Fermi energy in $Ta_2O_5$ and $TiO_2$, respectively. $Ta_2O_5$ supercell of 112 atoms and $TiO_2$ supercell of 72 atoms are built and one oxygen vacancy is created in each supercell. We study three types of Ov sites in λ $Ta_2O_5$ and one Ov site in rutile $TiO_2$. The in-cell atomic structures are relaxed by GGA functional. We apply sX-LDA functional to the relaxed structures to calculate the Ov formation energy. The charge state and cell size corrections to the defect formation energies are handled by the method of Lany and Zunger [55]. Due to considerable computational burden of the non-local functional calculation, only Gamma point is used to sample the brillouin zone, as is justified by the reasonably large cell size. As the charge transition levels do not depend on the oxygen chemical potential, we set the oxygen chemical potential to be one half of the free energy of $O_2$ molecule, which implies the oxygen rich condition. The formation energy diagrams are shown in Fig. 4. For $Ta_2O_5$, Ov at the two-fold coordinated intralayer oxygen site has the lowest formation energies for all charge states, in agreement with the result of Guo et al [56]. The charge transition level (+2/0) and (0/-2) lies 0.2 eV below and 0.2 eV above the conduction band (CB) edge, respectively. For $TiO_2$, the charge transition level (+2/0) lies 0.3 eV below the CB edge, similar to recent GW result [57]; the (0/-2) coincides with the CB edge. As shown in the last part, the calculated Fermi energies of both $Ta_2O_5/TaO_2$ and $TiO_2/Ti_4O_7$ interfaces are above the CB edges of respective stoichiometric oxides which implies the formation of -2 charged Ov instead of +2 charged ones. The stability of -2 charged Ov should indicate different switching operation which is surprising since the RS model based on the +2 charged Ov agrees well with the reported experimental results. The controversy may be resolved by taking the work function tuning effect of the gate material into account. Metal like Pt as a typical high work function gate material contacts with $TaO_2$ and tunes the equivalent work function to higher value that the +2 charged state can be stabilized. The tunable work function by the bilayer metal structure has been demonstrated for MOSFETs [58]. Nevertheless, it is useful for the O vacancy to be in the positive charge state, to ensure controlled drift under the switching field. Experimentally, scavenger metal layer has been used to optimize the performance of RS memory cells [59, 60]. Guo et al [56] rationalized that scavenger metal of larger work function than the parent metal electrode lowers the Fermi energy to ensure Ov in its positively charged state. In the following part, we propose the use of dopants as another route for this purpose.

**The effects of dopants**

Doping has been used to optimize the RS properties in $Ta_2O_5$ [30-32] and $TiO_2$ [20,29]. Performance enhancements due to doping, however, needs significant clarification. For instance, the longer endurance, which was attributed to doping, was explained by the lowered Ov formation energy [29]. According to the compact electrical "hour glass" model [61], however, a lower Ov formation energy will increase the possibility of forming new vacancies during switching cycles, so the vacancy number over switching cycles fluctuates. This would shorten endurance instead. Thus, the role of dopants in the RS process clearly needs further investigation.

Apart from the lack of a comprehensive understanding of the bulk effects of dopants, to the best of our knowledge, there is no report on the interfacial effects of dopants on the RS process. In the last part, we have seen that the system Fermi energies for both $Ta_2O_5/TaO_2$ and $TiO_2/Ti_4O_7$ interfaces lie above the CB edge (or just slightly below according to [52]) of respective stoichiometric oxides, which suggests that negatively charged (or neutral according to [52]) Ov are stable. Guo et al [56] has studied the effect of scavenging metal on tuning the Fermi level of oxide/parent metal electrode interface. The idea was that scavenging metal with higher work function sets the effective system Fermi level to near the mid-gap where +2 charged Ov is stabilized. In this part, we propose the interfacial effects of dopants that can be used to shift the system Fermi energy and stabilize Ov in positive charge state.

For the optimized $Ta_2O_5/TaO_2$ interface structure, we replace symmetrically one interfacial Ta on the $Ta_2O_5$ side by Ti, Si, Al, respectively and one oxygen by N. We only optimize the in-cell atomic structure. For cation substitution, the atomic structures of each doped interface and the corresponding GGA band structures are shown in Fig. 5. It can be seen that the Fermi energy of the interface system shifts to below the CB edge of $Ta_2O_5$, leading to a schottky contact. In particular, Ti and Si induce comparable Fermi energy shift by 0.3 eV, while Al induces the largest Fermi energy shift by 0.7 eV. Notice that the element electronegativity decreases in the sequence of Si>Al>Ti>Ta. It has been pointed out that electronegativity difference between dopant and host atoms at the interface induces dipole, resulting in the shift of the band offset and EWF as a consequence in the $HfO_2/SiO_2$ heterostructure [39]. The direction of shift is linked to the sign of difference of electronegativity [40]. Our results tend to support this point. As Si, Al and Ti are all more electronegative (or less electropositive) than Ta, the replacement of an interfacial Ta on the $Ta_2O_5$ side by one of these atoms results in reversed Ta($TaO_2$ side)-O-D(dopant) dipole with the original Ta($TaO_2$ side)-O-Ta($Ta_2O_5$ side) one; consequently, the electrostatic potential in metallic $TaO_2$ is lowered and the system Fermi energy shifts downward. Though dopants can disperse in bulk $Ta_2O_5$ and are not restricted to interfacial region, dopant induced dipole is much more effective at the interface than in the bulk due to the strong screening in high-K $Ta_2O_5$ [40].

Despite of the fact that all Si, Al and Ti modulate the system Fermi energy in the same direction, the strength of the modulation seems not to vary only with the electronegativity. This can be seen from the largest Fermi energy shift by Al regardless of its mediate electronegativity among these selected dopants. A possible explanation is by taking the valence states of dopants into consideration. Si and Ti are tetravalent, Al is trivalent. Considering the pentavalence of Ta, Si and Ti substitutions leave a hole on the top of the VB edge of $Ta_2O_5$ and electron transfer from the Fermi level of metallic $TaO_2$ fills the hole, leaving a substitution state of -1 charge. An additional interfacial dipole due to the charge transfer is built up. It is noteworthy that this additional dipole is of the same direction as the electronegativity induced one for Si, Al and Ti dopants. Both of these two types of interfacial dipoles tend to shift the system Fermi energy downward below the CB edge of $Ta_2O_5$. In the case of Al substitution, it leaves state of -2 charge involving even more pronounced interfacial charge transfer which may account for the most significant Fermi energy modulation. Zhao et al [33] found that the bulk effect of dopant also follows a valence-electron-based rule: both p-type and n-type dopants decrease the Ov formation energy. The valence dependent rule as pointed out here, however, clearly makes the distinction between p-type and n-type dopants.

An Ov model is also frequently invoked to explain the n-type flat band (FB) voltage shift in metal-oxide-semiconductor (MOS) structures [62,63]. The idea is that negatively charged interfacial substitution states tend to give rise to the formation of a positively charged Ov nearby or in the bulk in order to conserve the charge neutrality; and if the centroids of negative and positive charges do not coincide, a net dipole is created that modulates the system Fermi energy. We take the Ov model into consideration since Ov are abundant in RRAM and memristor devices and they are presumably to be in positive charge state. We use the Si doped interface model structure and create one Ov in the middle of bulk $Ta_2O_5$, the charge neutrality is therefore preserved considering two symmetrical Si substitutions and one eliminated O ($2Si_{Ta}^{-1}$ - $O^{2-}$). Ov in this case is effectively in the +2 state, forming a closed shell structure [64] by filling the holes created by $Si_{Ta}$ on top of the VB. This does have effect on the position of system Fermi energy. In particular, the net dipole created by $Si_{Ta}$-Ov pair is in the converse direction to the aforementioned interfacial dipoles, namely, electronegativity induced and interfacial charge transfer induced dipoles. As expected, we observe upward shift of Fermi level lying at the CB edge of $Ta_2O_5$ due to Ov, as shown in Fig. 6(b). The schematics of these three dipole effects are shown in Fig. 7. The strength of each of these three dipoles is subject to further study.

For anion substitution, there are two opposing dipoles associated with Ta on both sides of $Ta_2O_5$ and $TaO_2$, namely, O-Ta-N on $Ta_2O_5$ side and N-Ta-O on $TaO_2$ side. Considering the smaller electronegativity of N than O and the metallic screening of $TaO_2$, O-Ta-N dipole on $Ta_2O_5$ side effectively lifts up the electrostatic potential in $Ta_2O_5$ and results in downward shift of system Fermi energy to below the CB edge of $Ta_2O_5$, as shown in Fig. 8. This schematic of this dipole effect is illustrated in Fig. 7(d). Hinkle et al [65] reported interfacial nitrogen induced EWF increase in $HfO_2$/TiN stack, the rationale is the same.

Finally, we study a new element dopant, Li, at $TiO_2$/$Ti_4O_7$ interface. We replace symmetrically one interfacial Ti on $TiO_2$ side by Li. In contrast to cation substitutions at $Ta_2O_5$/$TaO_2$ as shown above, the substitutional Li is much less electronegative than the host Ti. Besides, Li is monovalent and Ti is tetravalent. This results in a pair of opposing interfacial dipoles induced by electronegativity difference and charge transfer, respectively; the former tends to lift up the Fermi energy higher in the CB of $TiO_2$, while the latter tends to lower the Fermi energy to below the CB edge. The charge transfer is so strong due to large difference of valence states that the competitive dipole is compensated and the system Fermi energy is shifted below the CB edge of $TiO_2$.

**Conclusion**

In this work, we use first principle calculation to study the interfacial properties of $TiO_2/Ti_4O_7$ and $Ta_2O_5/TaO_2$ interfaces. We find that both interfaces form insulator/metal like ohmic contacts, with the system Fermi energies above the CB edges of respective stoichiometric oxides. This leads to important implications that Ov are stabilized in the negatively charged state, considering the dependence of charge state stability of Ov on the position of Fermi level with respect to the band edge in $TiO_2$ and $Ta_2O_5$, respectively. We propose the use of interfacial dopants to modulate the system Fermi energy. Four dipole mechanisms have been identified induced by interfacial dopants and downward shift of system Fermi energies to below the CB edge of $TiO_2$ and $Ta_2O_5$, respectively, is observed by cooperative or competitive interplay among these dipole mechanisms. Further studies of $TiO_2/Ti_4O_7$ and $Ta_2O_5/TaO_2$ interfaces by higher level method, and the effectiveness of various interfacial dopants and dipole mechanisms are needed.

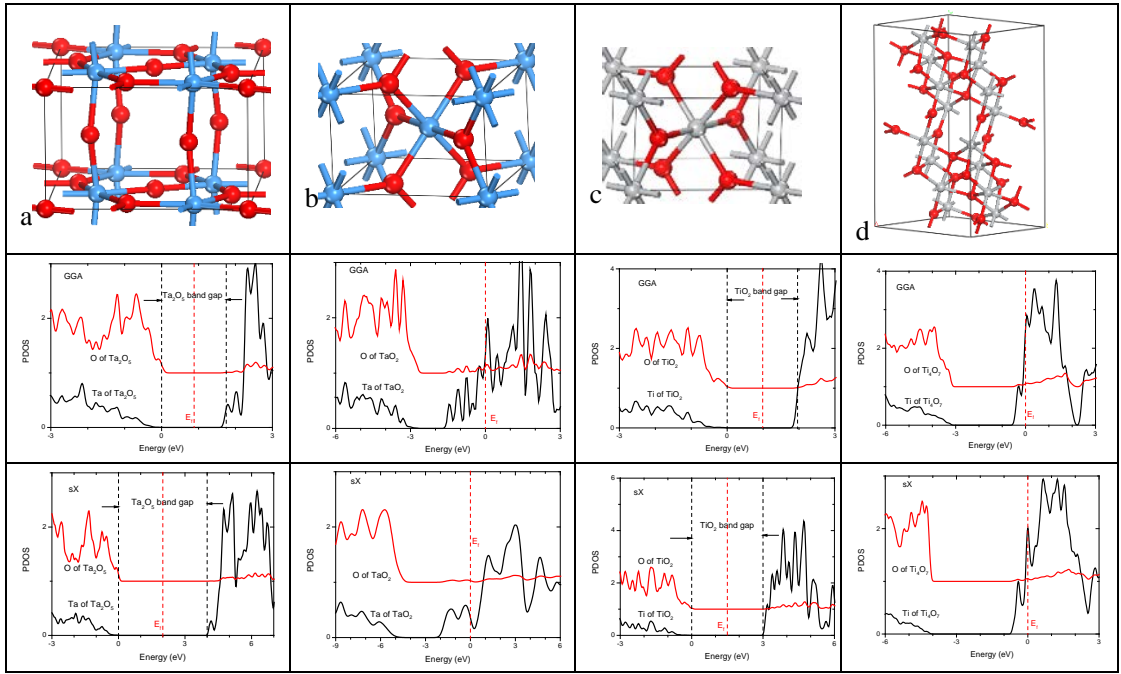

Fig. 1 The GGA relaxed crystalline structures (top) and the corresponding GGA (middle), sX-LDA (bottom) band structures of (a) λ $Ta_2O_5$, (b) rutile-$TaO_2$, (c) rutile-$TiO_2$, and (d) $Ti_4O_7$. The red balls represent for O, the blue ones for Ta and the grey ones for Ti. The Fermi levels of each system are indicated by a red vertical dash line. For $Ta_2O_5$ and $TiO_2$, the VB and CB edges are indicated by black vertical dash lines.

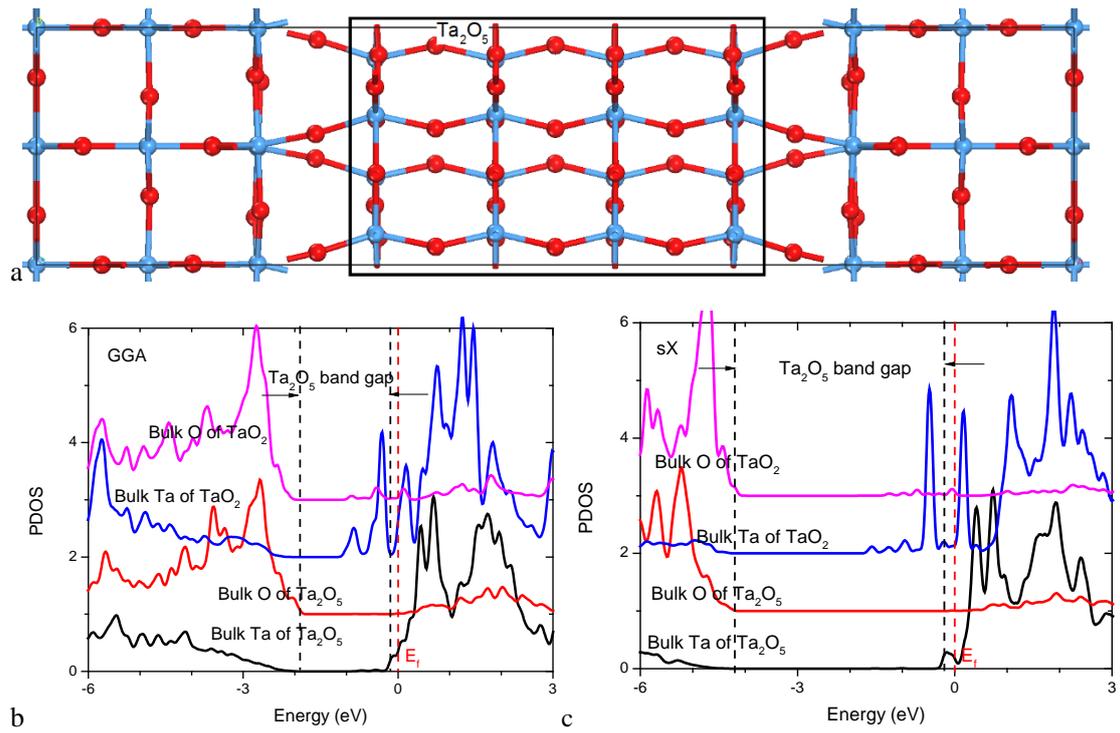

Fig. 2(a) The GGA relaxed atomic structure of $Ta_2O_5/TaO_2$ interface. The (b) GGA and (c) sX-LDA partial density of states (PDOSs) of representative atoms in the middle of bulk $Ta_2O_5$ and $TaO_2$.

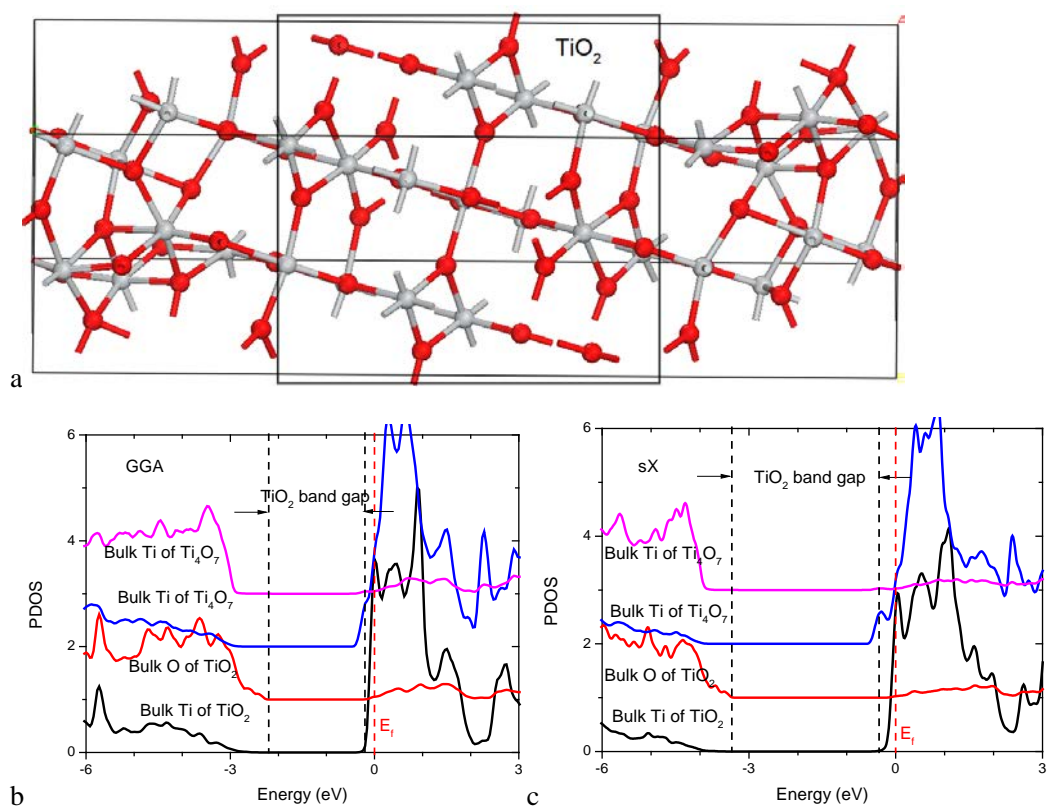

Fig. 3(a) The GGA relaxed atomic structure of TiO$_2$/Ti$_4$O$_7$ interface, The (b) GGA and (c) sX-LDA partial density of states (PDOSs) of representative atoms in the middle of bulk TiO$_2$ and Ti$_4$O$_7$.

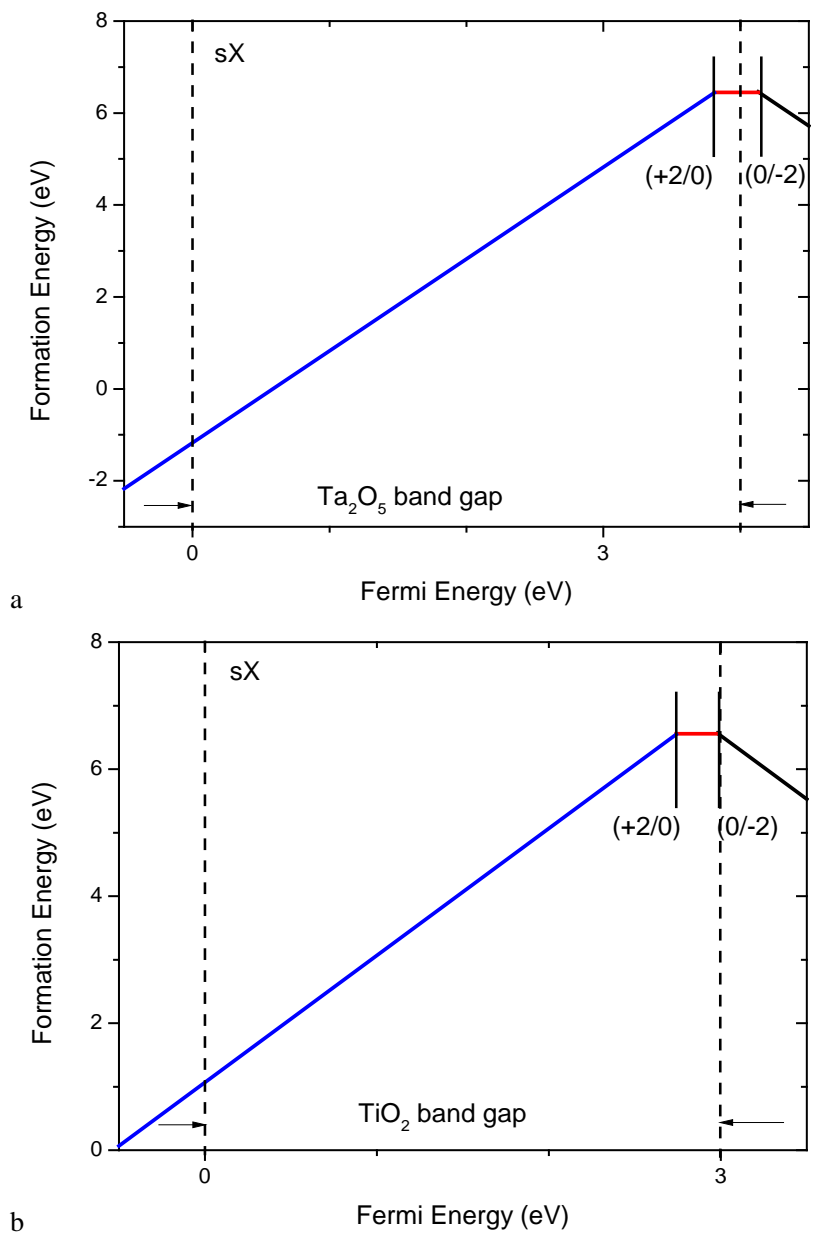

Fig. 4 Formation energy diagram of an Ov in λ Ta$_2$O$_5$ and rutile-TiO$_2$ calculated by sX-LDA. The band edges are indicated by black vertical dash lines.

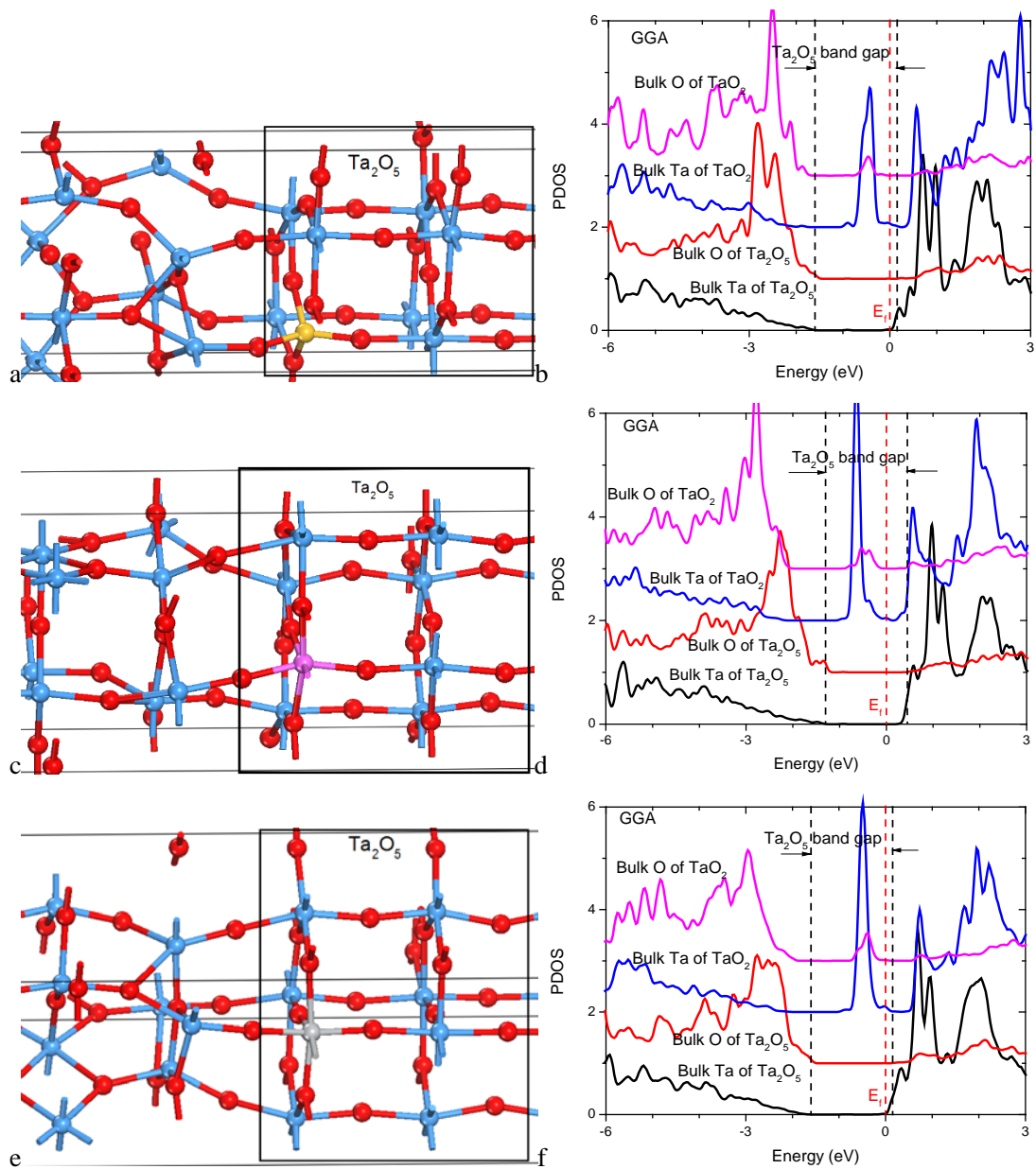

Fig. 5 The GGA relaxed atomic structures of interfacial cation substitutions on Ta$_2$O$_5$ side of Ta$_2$O$_5$/TaO$_2$ interface by (a) Si, (c) Al, and (e) Ti; the corresponding GGA PDOSs are shown in (b), (d) and (f), respectively. The golden atom represents for Si, the pink one for Al and the grey one for Ti.

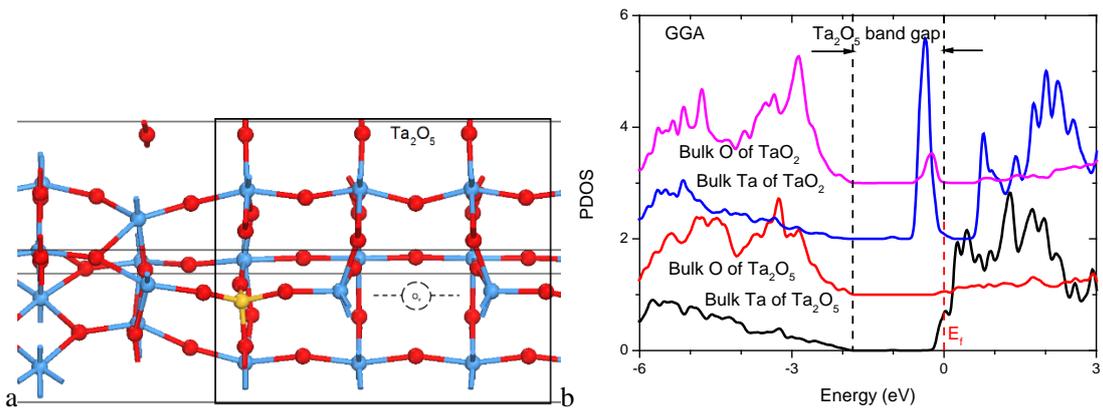

Fig. 6 (a) The GGA relaxed atomic structure of interfacial Si substitution on $Ta_2O_5$ side at $Ta_2O_5$/$TaO_2$ interface and a bulk Ov in $Ta_2O_5$, and (b) the GGA PDOSs.

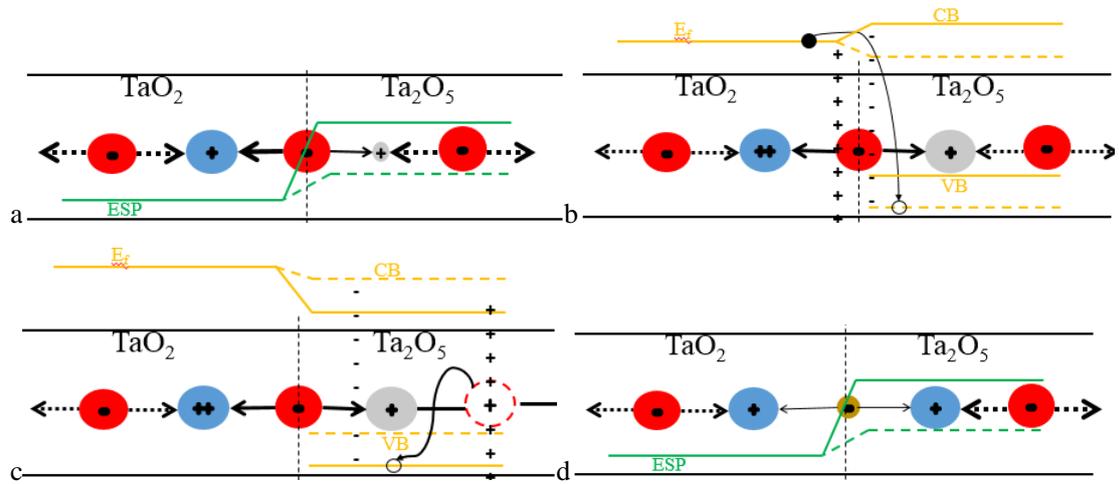

Fig. 7 Schematics of four interfacial dopant induced dipole effects in shifting downward the system Fermi energy, (a) substitutional more electronegative cation, indicated by small grey ball, for an interfacial Ta on $Ta_2O_5$ side; the resulting O-D(dopant) dipole is smaller, indicated by thin full arrow line; dipoles in bulk are screened, indicated by dotted arrow line; the dash and full green lines represent for electronic static potential (ESP) before and after dipole formation, respectively; (b) substitutional lower valence state cation, indicated by the grey ball labelled by only one "+", for an interfacial Ta on $Ta_2O_5$ side; a hole, indicated by open circle, is created at the top of $Ta_2O_5$ VB; charge transfer from the Fermi level ($E_f$) of $TaO_2$ to fill the hole occurs and a dipole is built up across the interface; the orange lines represent for $E_f$ and CB/VB of $TaO_2$ and $Ta_2O_5$, respectively; (c) the same as (b) except for an additional Ov, indicated by red dash-line circle, in bulk $Ta_2O_5$; charge transfer from the Ov to fill the hole occurs and a dipole is built up; (d) substitutional less electronegative anion, indicated by small brown ball, for an interfacial O; the resulting D(dopant)-Ta dipole is smaller, indicated by thin full arrow line; dipoles in bulk $TaO_2$, indicated by thin arrow line, is more effectively screened than in $Ta_2O_5$.

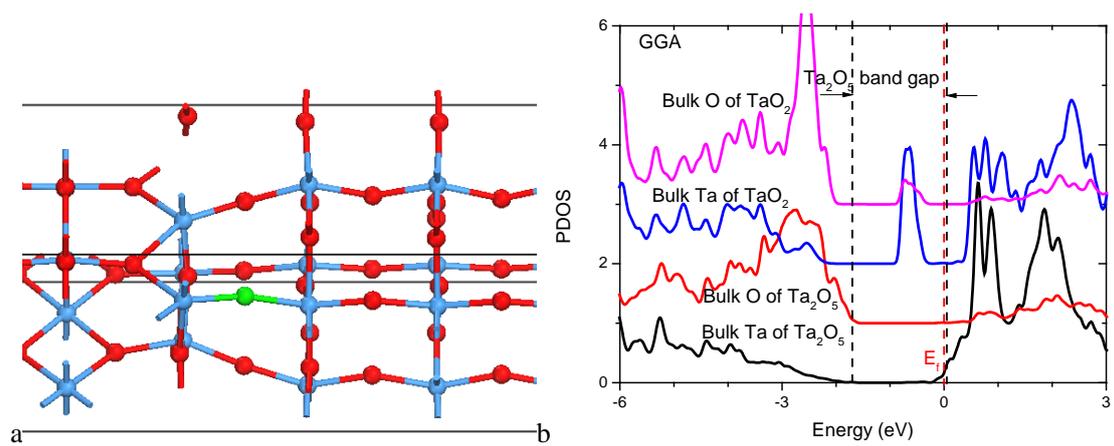

Fig. 8 (a) The GGA relaxed atomic structure of interfacial N substitution on $Ta_2O_5$ side at $Ta_2O_5/TaO_2$ interface and (b) the GGA PDOSs. The green atom represents for N.

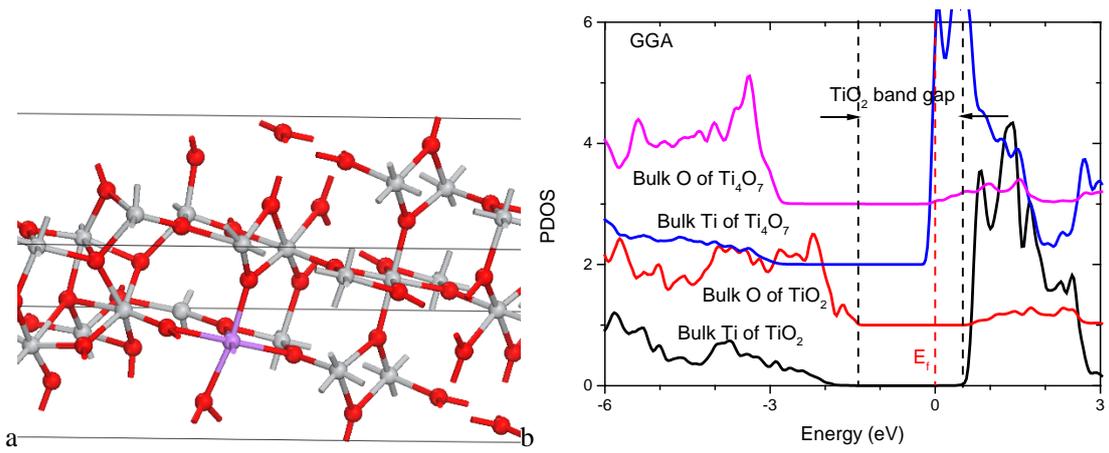

Fig. 9 (a) The GGA relaxed atomic structure of interfacial Li substitution on TiO$_2$ side at TiO$_2$/Ti$_4$O$_7$ interface and (b) the GGA PDOSs. The purple atom represents for Li.